\documentclass[reprint,amsmath,amssymb,aps,prd,showpacs]{revtex4-1} 
\usepackage{graphicx}
\usepackage{dcolumn}
\usepackage{bm}
\usepackage{url}

\newcommand{\de}[0]{\partial}

\begin{document}

\preprint{APS/123-QED}
\title{Effect of interference on thermal noise and coating optimization in dielectric mirrors}
\author{N. M. Kondratiev}
\author{A. G. Gurkovsky}
\author{M. L. Gorodetsky}
\email{gorm@hbar.phys.msu.ru}
\affiliation{Faculty of Physics, Moscow State University, Leninskie Gory, Moscow, 119991, Russia.}
\date{\today}

\begin{abstract}
Optical multilayer coatings of high-reflective mirrors significantly determine the properties of Fabry-Perot resonators. Thermal (Brownian) noise in these coatings produce excess phase noise which can seriously degrade the sensitivity of high-precision measurements with these cavities, in particular in laser gravitational-wave antennas (for example project LIGO), where at the current stage it is one of the main limiting factors. We present a method to calculate this effect accurately and analyze different strategies to diminish it by optimizing the coating.

Traditionally this noise is calculated as if the beam is reflected from the surface of the mirror fluctuating due to the sums of the fluctuations of each layer. However the beam in fact penetrates a coating and Brownian expansion of the layers leads to dephasing of interference in the coating and consequently to additional change in reflected phase. Fluctuations in the thickness of a layer change the strain in the medium and hence due to photoelastic effect change the refractive index of this layer. This additional effect should be also considered. It is possible to make the noise smaller preserving the reflectivity by changing the total number of layers and thicknesses of high and low refractive ones. We show how this optimized coating may be constructed analytically rather then numerically as before. We also check the possibility to use internal resonant layers  and optimized cap layer to decrease the thermal noise.
\end{abstract}


\maketitle

\section{Introduction}

Any precise measurement faces a challenge of different noises superposing useful signal. Brownian noise coming from chaotic thermal motion of particles is one of the examples. The Michelson interferometer is able to detect small changes in the lengths of its arms: two beams traveling different optical paths, interfere on the detector producing intensity which depends on the difference between phases of the beams. Thermal (sometimes also called Brownian) noise in coatings and substrates of the mirrors of the interferometer results in fluctuations of mirrors' surfaces averaged over the beam spot and adds a random phase to the waves. This effect is one of the key factors limiting the sensitivity of laser gravitational wave detectors \cite{LIGO}. Though the thickness of the multilayer coating is just several micrometers, the internal mechanical losses in layers is several orders of magnitude larger then in the substrate. That's why coating thermal noise, in accordance with the fluctuation-dissipation theorem, dominates and exceeds  other noises produced in the mirrors \cite{Zoo}.

In this paper we analyze different effects and strategies aimed to decrease the thermal coating noise for
generalized multilayer reflective coating. Traditionally this noise is calculated as if the optical beam is reflected from the surface of the mirror fluctuating as the incoherent sum of the fluctuations of each layer. However, the beam actually penetrates the coating and Brownian expansion of the layers leads to dephasing of interference and consequently to additional change in reflected phase \cite{Gurkovsky}. Fluctuations in the thickness of a layer change the strain in the medium and hence due to photoelastic effect change the refractive index of this layer. This additional effect should also be considered. It is possible to make the noise smaller while preserving the reflectivity by changing the number of layers and thicknesses of high and low refractive components \cite{Akira,Harry}. We also check the possibility of using internal resonant layers \cite{Kimble} and an optimized cap layer \cite{Zoo} to decrease the thermal noise.

Brownian noise is not the only source of noise produced by the coating. Fluctuations of temperature, which translate into displacement of mirror's surface through thermal expansion (thermoelastic noise) \cite{BGVte,BGVteC} and change of optical path due to fluctuations of refraction index (thermorefractive noise) \cite{BGVtr} combine producing generalized thermo-optical noise \cite{Zoo,Evans}. Brownian fluctuations causing displacement of mirrors' surface and the previously neglected correlated photoelastic effect produced by these fluctuations both lead to fluctuations of the optical path and should be treated simultaneously. The Brownian branch of noises, which is the topic of this paper, and thermo-optic one are uncorrelated as they represent uncorrelated fluctuations of volume and temperature.

\section{Phase noise from multilayer coatings}
\label{source}

\subsection{Reflectivity}

To calculate the amplitude and phase of a reflected beam the impedance method \cite{Haus} will be used below. We found this method more convenient for analytical consideration than the equivalent and more widely used  matrix method \cite{Tikhonravov}.

\begin{figure}[b]
\center
\includegraphics[width=0.45\textwidth]{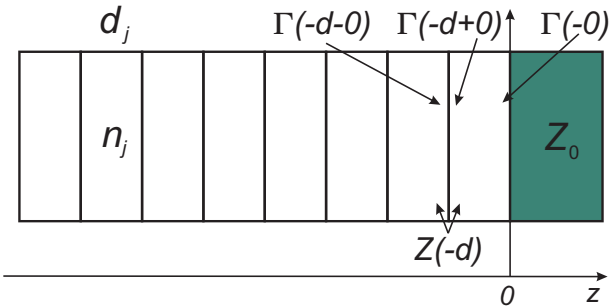}
\caption{A schematic of a multilayer coating}
\label{impmp}
\end{figure}

To consider reflection on each boundary separating the layers starting from the substrate/coating boundary (see pic. \ref{impmp}) we introduce an effective impedance $Z(z)$ and amplitude reflection coefficient $\Gamma(z)$ as follows:
\begin{align}
\label{Impm1}
Z(z)&=\frac{E(z)}{H(z)}=\frac{E_+(z)+E_-(z)}{H_+(z)+H_-(z)}=\eta(z)\frac{1+\Gamma(z)}{1-\Gamma(z)},\\
\label{Z->G}
\Gamma(z)&=\frac{E_-(z)}{E_+(z)}=\frac{Z(z)-\eta(z)}{Z(z)+\eta(z)},\\
\eta(z)&=\sqrt{\frac{\mu(z)\mu_0}{\epsilon(z)\epsilon_0}}=\frac{\mu(z)}{n(z)}Z_v.
\end{align}
where $E$ and $H$ are tangential electric and magnetic fields in the standing wave, while $E_+$, $H_+$ and $E_-$, $H_-$ are forward and backward (reflected) waves, $n$ is the refraction index, $\mu$ and $\epsilon$ the relative permeability and permittivity, and $Z_v$ is the vacuum impedance ($Z_v=1$ in Gaussian CGS system). 

As tangential fields $E$ and $H$ are continuous in a medium without free currents,  the effective impedance is also continuous on all  boundaries, while reflection coefficient experiences jump. Meanwhile the reflection coefficient changes continuously between boundaries according to the following equation:
\begin{align}
\label{propagator-1}
\Gamma(z-d_j)&=\frac{E_-e^{i\omega t+ik_0n_j(z-d_j)}}{E_+e^{i\omega t-ik_0n_j(z-d_j)}}=\Gamma(z)e^{-i2k_0n_jd_j},
\end{align}
where $k_0=\frac{2\pi}{\lambda}$ is the wave vector of the optical field in vacuum, $\lambda$ is the wavelength. This allows one to calculate any multilayer coating layer by layer recursively starting from the substrate in which the impedance is equal to the impedance of free substrate $\eta_s$ and moving to the surface, turning from the reflection coefficient $\Gamma_j=\Gamma(-\sum_j d_j-0)$ to the effective impedance $Z_j=Z(-\sum d_j -0)=Z(-\sum d_j +0)$ when facing the boundary and back after crossing it (see Fig. \ref{impmp}). In particular it is possible to exclude effective impedance from calculations by deriving the formula for reflectance jump itself:
\begin{align}
\label{jump}
\Gamma_{j+1}&=\frac{g_{j+1,j}+\Gamma_je^{-i\varphi_j}}{1+g_{j+1,j}\Gamma_je^{-i\varphi_j}},
\end{align}
where $g_{ij}=\frac{n_i-n_j}{n_i+n_j}$, $\varphi_{j}=2k_0n_jd_j$.

In the case of classical reflective $\lambda/4$ layers, all impedances and reflection coefficients are real.

\subsection{Interference}

We now assume that each of the layers can have variation in thickness $\delta d_j$ and independent variations of its refraction index $\delta n_j$ arising from fluctuations. Both should be accounted in \eqref{propagator-1}, introducing $-\varphi_j\rightarrow-\varphi_j-2k_0\delta n_jd_j-2k_0n_j\delta d_j=-\varphi_j+\Delta_j$. We also have to rewrite $\eta_i\rightarrow\eta_i(1+\delta \eta_i)$ in \eqref{Impm1}-\eqref{Z->G}, which is a consequence of refraction index change $\delta \eta_j=-\frac{\delta n_j}{n_j}$. As before, moving layer by layer to the surface, we expand each result into a series to the first order of variations $\delta n_j$ and $\delta d_j$. In this way we can build a perturbed amplitude reflection coefficient $\Gamma'_m$:
\begin{align}
\label{mainFormulas}
\Gamma'_m&=\Gamma_m(1+\varepsilon),\nonumber\\
\varepsilon&=f_m\frac{\delta n_m}{n_m}+\sum_{j=1}^{m-1}\!\prod_{k=j+1}^m \!f_kz_{k-1}\left(i\Delta_j-\mu_j\frac{\delta n_j}{n_j}\right),
\end{align}
\begin{align}
\label{mainKoefs}
f_k&=\frac{(1-\Gamma_{k}^2)}{2\Gamma_{k}},\ 
z_k=\frac{2\Gamma_ke^{-i\varphi_k}}{1-\Gamma_k^2e^{-i2\varphi_k}},\ 
\mu_k=\frac{1}{z_k}-f_k.
\end{align}
Here $m$ is the index of the layer of interest ($m=N+e$ for total reflectance, where $N$ is the full number of layers, ``e'' in indices represents external medium and ``$+e$'' means ``$+1$'') Taking into account that $\Delta_j,\frac{\delta n_j}{n_j}\ll 1$, we can find the equivalent phase shift $\delta\varphi$ as well as modification of reflectivity $\delta\Gamma$ (leading to amplitude noise which cannot be found in traditional approach) collecting all imaginary and real parts noting the decomposition: $\Gamma e^{\varepsilon}=\Gamma(1+\varepsilon)$.
Total fluctuations may arise both from layer thickness fluctuations $\delta d_j$ (Brownian and thermoelastic noises), or from deviations of refraction index $\delta n_j$ (photoelastic and thermo-refractive noises). 

\subsection{Photoelastic effect}

Besides surface displacement noise there should be a photoelastic noise in the Brownian noise branch. Photoelasticity is a phenomenon of refraction index change under deformation:
\begin{align}
\Delta B_i=p_{ij}u_j,
\end{align}
where $B_i$ is the optical indicatrix, $u_j$ -- strain tensor, $p_{ij}$ is photoelastic tensor and indices $i,j\in 1;6$ \cite{Acoustoopt}. In case of cylindrical symmetry we have longitudinal effect $\Delta B_i=p_{i3}u_3=p_{i3}\delta d/d$ and transversal effect $\Delta B_i=p_{i\rho}u_{\rho\rho}$. However, only noise produced by longitudinal effect has the same origin as thermal one we consider (movement in ``z'' direction), and in this way there is a theoretical possibility of their interference compensation. For longitudinal effect of variation of refraction indices one obtains:
\begin{align}
\delta n_x=-\frac{n_0^3}{2}p_{13}\frac{\delta d}{d},\nonumber\\
\delta n_y=-\frac{n_0^3}{2}p_{23}\frac{\delta d}{d}.
\end{align}
There is also nonzero $\delta n_z$ component, but we consider only rays incoming perpendicularly to the surface and do not take it into account. It is known that tantalum oxide used in multilayer coatings $Ta_2O_5$ -- is a rutile (titanium oxide) type crystal with tetragonal symmetry. Rutile has $p_{13}=0.171$, $p_{23}=0.16$. From \cite{Tantal} we can make a rough estimation for tantalum oxide $p_{Ta_2O_5}<0.18$. We put for simplicity $p_{13}=p_{23}=p_{Ta_2O_5}=0.17$. The other component of the coating -- fused silica has $p_{13}=p_{23}=p_{SiO_2}=0.27$. 

The transversal effect should be considered separately as $u_{\rho\rho}$ -- motion is not correlated with $u_{zz}\propto\delta d$ -- motion. This effect should not give out more noise than the longitudinal one and should be added incoherently.

\subsection{Brownian branch of noises}

We use longitudinal photoelastic effect to convert fluctuations of refraction index into fluctuations thickness of layer: 
\begin{align}
\Delta_j&=-2k_0n_j\left(1-\frac{n_j^2}{2}p_{j}\right)\delta d_j=-2k_0n_j\psi_j\delta d_j,\\
-\frac{\delta n_j}{n_j}&=\frac{n_j^2p_j}{2}\frac{\delta d_j}{d_j}= -\frac{n_j^2p_j}{\varphi_j(2-n_j^2p_j)}\Delta_j=\gamma_j\Delta_j,
\end{align}
where $p_j$ is effective photoelastic index for $j$-layer. Then coating induced deviations of reflected phase and refletion coefficient are
\begin{align}
\label{BrownNoiseInterf}
\delta\varphi_{c}&=\sum_{j=1}^{N}\alpha_j\delta d_j,\\
\label{BrownNoiseAmpl}
\delta\Gamma_{c}&=\sum_{j=1}^N\beta_j\delta d_j,
\end{align}
where
\begin{align}
\alpha_j=&-2k_0n_j\psi_j\Im[\prod_k f_kz_{k-1}(i+\mu_j\gamma_j)],\\
\beta_j=&-2k_0n_j\psi_j\Re[\prod_k f_kz_{k-1}(i+\mu_j\gamma_j)].
\end{align}

\begin{figure}
\center
\includegraphics[width=0.4\textwidth]{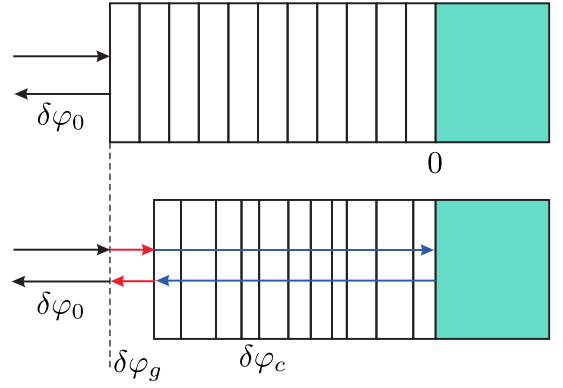}
\caption{Phase shift of the optical wave reflecting from an unperturbed (upper figure) and perturbed mirror. $\delta\varphi_0$, $\delta\varphi_g$ and $\delta\varphi_c$ are shift in the total phase, the shift due to the surface displacement and the shift due to interference dephasing in the coating respectively ($\delta d_j<0$).}
\label{NoiseIllus}
\end{figure}

Let us consider one end mirror in the arm of an interferometer. Thermal displacement of the mirror's surface produces phase fluctuations in the interferometer output. It is more intuitive to consider the case of contraction (Fig. \ref{NoiseIllus}) of the mirror. Then the length of additional gap for the light to travel before entering the mirror is $-\delta d$ (as $\delta d<0$ for contracting), yielding the phase shift
\begin{align}
\label{BrownNoiseTrad}
\delta\varphi_g&=-2k_0\sum_{j=1}^{N}(-\delta d_j),
\end{align}
The total phase shift produced by the perturbed coating (relative to the unperturbed one) will be
\begin{align}
\label{BrownNoiseFull}
\delta\varphi_{\Sigma}&=-2k_0\sum_{j=1}^{N}\left[f_{N+e}(-1)^{N-j} z_j\psi_jn_j-1\right]\delta d_j,
\end{align}
where we took into account that inside $\lambda/4$-reflector all quantities are real and $\alpha_j=-2k_0n_j\psi_jf_{N+e}(-1)^{N-j} z_j$.

It is also important to admit that in a ``good mirror'' approximation, when $1-|\Gamma|\ll 1$ (in this case depending on the topmost layer $Z_N\rightarrow0$ or $Z_N\rightarrow\infty$) the amplitude reflection coefficient correction produced by each layer $\beta_j=(-1)^{N-j}f_{N+c}\gamma_jf_j\frac{Z_j}{\eta_j}\rightarrow0$ (for $\lambda/4$-reflector).

The term before $\delta d_j$ can be treated as a noise coefficient showing a contribution of each layer into the total noise. This coefficient can have any sign, depending on the values of interferential contribution (``$-$'' sign) or surface displacement (``$+$''), but only its absolute value is significant as noise contributions from different layers are added incoherently.
\begin{figure}
\center
\includegraphics[width=0.45\textwidth]{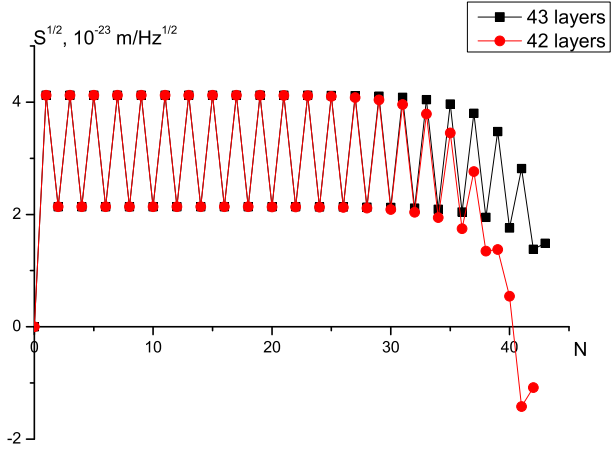}
\caption{Noise coefficient (keeping the sign) from each layer in a coating consisting of 42 (circles) or 43 (squares) layers on silicon substrate.}
\label{AlphaShumSiPe}
\end{figure}

Using the acquired formulas we can plot a diagram of phase shift contribution of each layer and values of noise spectral densities in the whole. In Fig. \ref{AlphaShumSiPe} such distribution is plotted, keeping the sign from (\ref{BrownNoiseFull}). It can be seen that the  interference part of noise plays a role in a few outer layers (order of penetration depth) \cite{Gurkovsky} while Brownian (surface displacement) noise forms the major part. Several layers can even demonstrate nearly complete noise compensation.

We should note that the noise contribution of a layer is formally composed of three summands: main Brownian (surface displacement) contribution, interferential part and photoelastic effect:
\begin{align}
\label{DerivativeMethod}
\delta \varphi_{\Sigma}=\sum 2k_0\delta d_j+\frac{\de \varphi}{\de d_j}\delta d_j+\frac{\de \varphi}{\de n_j}\frac{\de n_j}{\de d_j}\delta d_j,
\end{align}
where $\varphi$ denotes the phase of complex reflectivity. Formulas \eqref{mainFormulas}-\eqref{mainKoefs} give analytical expression of the derivatives. Their sign distribution may be illustrated as follows. If the coating contracts, then the phase shift produced by each layer is positive due to the change of its thickness as the Brownian (surface displacement) noise is not really a phase shift acquired by light inside the mirror, but outside it (Fig. \ref{NoiseIllus}). The contraction of each layer at the same time leads to an increase of refraction index as in normal materials it grows with density providing positive phase shift. Interference dephasing (phase shift due to reduction of layer thickness itself), on the other hand, is somewhat that compensates the phase shift produced by both effects. It looks like photoelastic effect can play only negative role, however it can be the thing to reduce too high interference dephasing (for example it lowers the total noise in 42-layer $\lambda/4$-coating).

Equation \eqref{DerivativeMethod} is quite easy for numerical calculations as the partial derivatives in it may also be calculated numerically. We used this approach for independent checking of formulas \eqref{mainFormulas}-\eqref{mainKoefs}.

\subsection{Noise spectral density}

Using \eqref{BrownNoiseFull} one can estimate noise spectral densities if the noise spectral density of each layer is known. In the model of independent thin layers on an infinite half space substrate, each layer behaves just as if it was the only layer on the substrate. This model was heavily treated and the solution is well known \cite{Harry,Gurkovsky}. For our purpose however, we should split the total surface fluctuations of one layer into two parts. The first one being the fluctuations of the thickness of the coating layer $S^c$ and the second one being the fluctuations of the substrate surface induced by losses in the coating $S^s$. Interference and photoelastic effects influence only the first term. If the losses in the layer responsible for both fluctuations (shear and expansion losses) are equal, what we assume in this paper, then these two spectral densities are uncorrelated. Otherwise cross correlation terms should be taken into account.
This splitting may be easily obtained using the approach presented in \cite{Gurkovsky}, assuming that the noise produced by each layer is independent $\langle\delta d_j^2\rangle\rightarrow S^c(\Omega)_j$, $\langle\delta d_j\delta d_k\rangle=0$:
\begin{align}
S(\Omega)_{j}&=S^c(\Omega)_{j}+S^s(\Omega)_{j}=(\xi^c_j+\xi^s_j)\phi_j d_j=\xi_j\phi_j d_j, \label{sigmochka2}\\
\xi^c_j&=\frac{4k_BT}{\pi w^2\Omega}\frac{(1+\nu_j)(1-2\nu_j)}{Y_j(1-\nu_j)}
,\nonumber\\
\xi^s_j&=\frac{4k_BT}{\pi w^2\Omega}\frac{Y_j(1+\nu_s)^2(1-2\nu_s)^2}{Y_s^2(1-\nu_j^2)}
\end{align}
where $\nu_j$ is the Poisson coefficient of layer $j$, $Y_j$ -- its Young's modulus ($Y_s$ and $\nu_s$ are the  parameters of the substrate), $\phi_j$ is the mechanical loss angle,  $w$ is the Gaussian beam radius on the mirror, $\Omega$ is frequency of analysis, $k_B$ is the Boltzmann's constant and $T$ is the temperature. Then we get spectral densities of phase and amplitude reflection fluctuations
\begin{align}
S_{\varphi}=&4k_0^2\sum_{j=1}^{N}[\left(\alpha_j-1\right)^2 S^c(\Omega)_{j}+S^s(\Omega)_j],\\
S_{\Gamma}=&4k_0^2\sum_{j=1}^{N}\beta_j^2 S^c(\Omega)_{j}.
\end{align}
The first sum can be simplified and the second is zero in assumption of a ``good mirror'' and $\lambda/4$ layers:
\begin{align}
S_{\varphi}=&4k_0^2\sum_{m=1}^2[S^c(\Omega)_{m}\left(\frac{a_m^2\psi_m^2}{|n_1^4-n_2^4|}-\frac{2a_m\psi_m}{|n_1^2-n_2^2|}+N\right)\nonumber\\
&+S^s(\Omega)_{m}N]
\end{align}
for $2N$ layers, and $S_{\varphi}+4k_0^2S_{1}$ for $2N+1$ layers, where 
$a_m=n_m^2$ for zero outer impedance ($2N$ layers with $n_1>n_2$) and 
$a_m=n_2^2n_1^2$ for infinite outer impedance ($2N+1$ layers).

To simplify the comparison of this type of noise with other types of noises and Fabry-Perot coordinate sensitivity we turn phase noise into noise of effective reflecting surface displacement
\begin{align}
S_x=&\frac{S_{\varphi}}{4k_0^2},
\end{align}
(in units of ${\text m}^2/\text{Hz}$) at 100 Hz frequency.

Calculations were made for a silica-tantala mirror of 42-43 layers (21 pairs of $SiO_2$ $Ta_2O_5$ $\lambda/4$-layers on fused silica substrate with or without additional $\lambda/4$-layer).
\begin{align}
\nu_l&=0.17,& \ &n_l=1.45, \nonumber\\
\nu_h&=0.23,& \ &n_h=2.06, \nonumber\\ 
Y_l&=7.2\times10^{10} \text{Pa},& \ &\phi_l=0.4\times10^{-4},\nonumber\\
Y_h&= 14\times10^{10} \text{Pa},& \ &\phi_h=2.3\times10^{-4}.\nonumber
\end{align}
\begin{align}
\lambda=1.064\times10^{-6} \text{ m};\ w=0.06 \text{ m};\ T=290 \text{ K}\nonumber
\end{align}
Results were formed in table \ref{SiTaTab} in form of correction $\chi=\frac{\sqrt{S_{Br}}-\sqrt{S}}{\sqrt{S_{Br}}}\times100\%$.  Numerical estimations for relative transmittance noise is $\delta \tau/\tau=2\Gamma\sqrt{S_\Gamma}/(1-|\Gamma|^2)<10^{-12}$ $\text{Hz}^{-1/2}$.

\begin{table}
\begin{center}
\begin{tabular}{|c|c|c|c|}
\hline
Type & $42\times\lambda/4$ &$41\times\lambda/4+ \lambda/2$ & $43\times\lambda/4$ \\\hline
Transmittance $\tau$, ppm & $2.28$ & $1.08$ & $0.54$  \\\hline
\hline
Brownian $10^{-20}$m$/\sqrt \text{Hz}$	& $0.632$ & $0.635$& $0.645$  \\\hline
$\chi$ With interference & $1.96\%$ & $2.34\%$ & $1.75\%$ \\\hline
$\chi$ With photoelasticity & $2.33\%$ & $1.85\%$ & $1.31\%$ \\\hline
$\chi$ Modified cap & $2.33\%$ & $2.76\%$ & $1.81\%$ \\\hline
\end{tabular}
\caption{\label{SiTaTab}Silica-tantala mirror efficiencies relative to the Brownian noise. Standard LIGO coating consists of 41 Layers$+\lambda/2$ cap mirror. Modified cap has optical width $\lambda/4$ (42 layers case).}
\end{center}

\end{table}

The interference correction to thermal coating thickness noise is about $6\%$, or $7.5 \%$ taking photoelasticity into account. But thickness variation of tantala layer is much smaller than its bending ($\xi^c_h=0.36\xi^s_h$). That is why the interference correction to full coating Brownian (displacement) noise is only about $2.0\%$, or $2.3 \%$ taking photoelasticity into account.

\section{Optimization strategies}
\subsection{Additional top layer-corrector}

One may tweak the thickness of the topmost ``correcting'' layer in an attempt to minimize the noises using interference effects. This method proved to be useful for thermoelastic and thermorefractive noises \cite{Zoo}. Using formulas \eqref{mainFormulas}-\eqref{mainKoefs} we can obtain
\begin{align}
S_{\varphi}=&4k_0^2\sum_{m=1}^2[S^c(\Omega)_m\left(\frac{a_m^2\psi_m^2}{|n_1^4-n_2^4|}-\frac{2a_m\psi_m}{|n_1^2-n_2^2|}+N\right)\nonumber\\
&+S^s(\Omega)_mN]+S^\prime_c\nonumber\\
S^\prime_c=&\left[\Re(g_{N+c+e})(1\pm\gamma_c\sin(\phi_c)) n_c\psi_c   -n_e\right]^2S^c(\Omega)_c\nonumber\\
&+S^s(\Omega)_c
\end{align}
for $2N$ layers, and $S_{\varphi}+4k_0^2S_1$ for $2N+1$ layers, where $a_m=n_m^2n_c\Re(g_{2N+c+e})$ and ``$+$'' for zero impedance of last but one layer ($2N+c$ layers with $n_2<n_1$ and $a_m= \frac{n_2^2n_1^2}{n_c}\Re(g_{2N+1+c+e})$ and ``$-$'' for infinite impedance of last but one layer ($2N+1+c$ layers). The index ``$c$'' represents the cap layer corrector and ``$+c$'' means ``$+1$''.
 
Results are quite unfavorable: for even number of layers + cap minimum of noise is at $n_c<1$ while its suppression $\chi=\frac{\delta \sqrt{S}}{\sqrt{S_{unmod}}}\times100\%$ is only $0.04\%$. For odd layers + cap absolute value of noise doesn't become lower than $6.198\times10^{-20}$ m$/\sqrt{Hz}$, which means suppression is less than $0.69\%$ (for $n_c=3.6; \ d_c=0.42\lambda/4$). Even after removing a pair of layers, the noise is about $6.04\times10^{-20}$ m$/\sqrt{Hz}$, which is more than for even number of layers.

This means that standard coating with top silica $\lambda/2$ layer is reflectance-optimized as well as ``all $\lambda/4$'' coating (cap$=\lambda/4$) is noise-optimized (see Tab. 1). 

\subsection{Layer-corrector inside the mirror.}
\begin{figure}
\center
\includegraphics[width=0.47\textwidth]{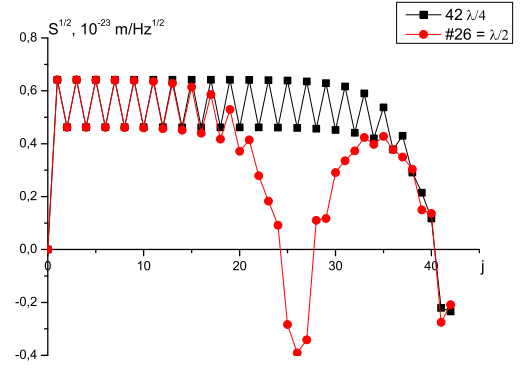}
\caption{Noise coefficient distribution in coating for 42 layers, keeping the sign. Silica substrate, vacuum medium (squares -- simple mirror; circles -- a mirror with modified layer \#26 (16 from top) $d_m=0.98\lambda/2$). }
\label{minimize1}
\end{figure}
In J.Kimble\cite{Kimble} proposed an idea of inserting a resonant layer into the mirror. This case was studied numerically (Fig. \ref{minimize1}). The maximum suppression of 4.4\% was shown by layer-corrector close to $d=\lambda/2$, which is the resonant cavity. But such modification increases power transmittance more than two orders of magnitude. If we add 8 bilayers to restore transmittance, suppression will be more than eliminated ($-14\%$).

\subsection{Two-sided and double mirror.}

\begin{figure}
\center
\includegraphics[width=0.47\textwidth]{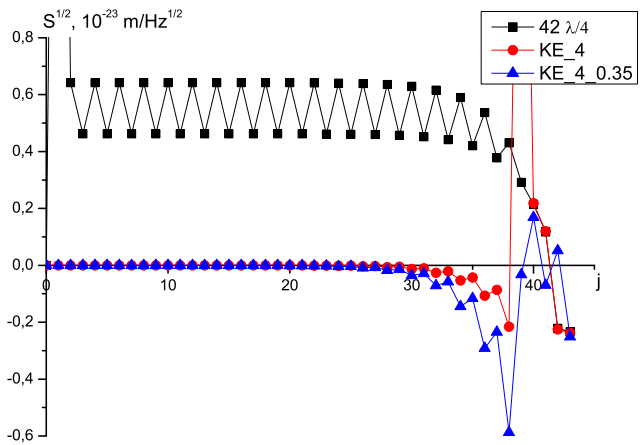}
\caption{Noise distribution in the coating, keeping the sign. Silica substrate, vacuum medium. Squares -- $\lambda/4$ general coating; circles -- corresponding etalon with $\lambda/4$ substrate; triangles -- etalon, optimized for interference, ($0.35\lambda/4$).}
\label{KeG}
\end{figure}

\begin{figure}
\center
\includegraphics[width=0.47\textwidth]{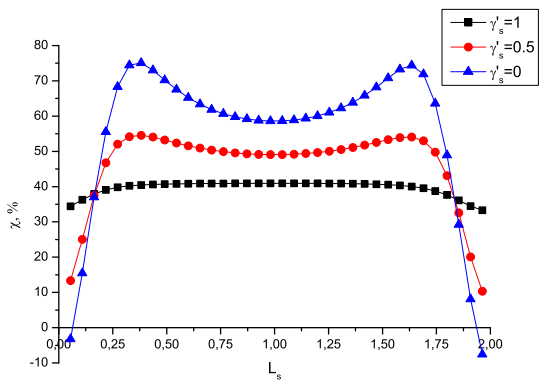}
\caption{Suppression as a function of excess substrate optical thickness (in units of $\lambda/4$) in addition to integer number of halfwavelengthes for different noise ratios $\gamma'_s=\frac{\xi^c_s}{\xi_s}$. }
\label{chi}
\end{figure}

 A novel combined structure was propose in \cite{Khalili} with just a few layers on the front side of a big silica substrate with antiresonance optical length and other layers moved to the the bottom (two-sided mirror or Khalili etalon). The idea is that only top layers produce Brownian noise, while bottom layers do not contribute to thermal noise as their surfaces move in the opposite direction. In this case we should also pay attention to interference effects, because first layers and substrate are well penetrated by light. That also means that coating noise and substrate noise in etalon should be treated simultaneously and there is a possibility of high interferential compensation (Fig. \ref{KeG},\ref{chi}).
The main difficulty with the Khalili etalon is its high sensitivity to the manufacturing precision and fluctuations of its optical thickness produced for example by temperature variations. Namely the imprecision of substrate optical length by $0.07\lambda/4$, corresponding to the mirror's temperature variation of 6 mK, increases noise by $5\%$.

The same idea may be realized in another geometry (double mirror or Khalili cavity) with combined end mirror consisting of two individually suspended mirrors separated by a controlled gap. The first mirror has small number of layers and hence low noise, while the layers of the second one provide the required reflectivity. The sensitivity to the gap length is two times higher, though it may be controlled with actuators in real time yielding desired conditions. Our calculations show encouraging suppression of noise in both schemes. The deficiency of both schemes however is high power circulating in the mirror's substrate, which leads to various thermal lensing and detuning effects.

The absolute value of maximum effect is highly dependent on ratio of thickness and bend noise spectral densities, which is yet unknown. This far we can only say that noise suppression and amplification effects decrease practically linearly with $\gamma'_s=\frac{\xi^c_s}{\xi_s}$ (Fig. \ref{chi}).

\subsection{Modifying silica-tantala ratio.}

A promising way to reduce the thermal noise in the coating was proposed \cite{Pinto,Akira}, which suggests decreasing the thickness of lossy high-index (tantalum-oxide) layers, presumably preserving the total bilayer optical thickness to be $\lambda/2$ ($n_ld_l+n_hd_h=\lambda/2$). To keep the required reflectivity, more bilayers should be used. It was found numerically that there is an optimum in the ratio of layers' thickness and number of layers providing minimal noise at a given reflectivity.

It appears that noise suppression $\chi$ is highly dependent on noise ratio in layers
\begin{align}
\chi\propto\frac{S_h/d_h}{S_l/d_l}=\frac{\xi_h\phi_h n_l}{\xi_l\phi_ln_h}=\gamma,
\end{align}
For the LIGO parameters \cite{GWINC} $\gamma=4.56$. In \cite{Akira} coating was optimized for a chosen parameter $\gamma=7$. An etalon silica-tantala mirror of 27 $\times\lambda/4$ layers+$\lambda/2$ cap was numerically optimized. Resulting coating had 16 silica-tantala bilayers with $n_ld_l=1.3827\lambda/4$, $n_hd_h=0.6173\lambda/4$, a thin cap $n_ld_l=0.1620\lambda/4$ and the first layer $n_hd_h=0.5560\lambda/4$ on substrate (34 layers total). In an experiment with this mirror design, noise suppression of $\chi_{exp}=(8.8\pm2.0)\%$ was observed. Our calculations with all material parameters taken from \cite{Akira} yield $\chi_{th_7}=8.2\%$, and if $\gamma=9.23$ estimated from the same experiment is used \cite{Krakow}, one gets $\chi_{th}=9.1\%$.

\section{Optimal coating}

It is well known that for a fixed number of bilayers, a multilayer coating with quarter length layers (QWL) with $\varphi_h=\varphi_l=\pi$  provides the largest reflectivity \cite{Tikhonravov}. The LIGO interferometers however require not only large reflectivity but also small noise added by the coating. The coating consists of two different materials having noticeably different mechanical losses. This fact stimulated mostly numerical attempts to construct more optimal coating which could have smaller noise with the increased number of layers but decreased total thickness of the ``bad'' component, while still preserving the desired reflectivity \cite{Akira,Pinto}. It was also found empirically that a better coating can be achieved when  $\varphi_h<\pi$ and it is a common knowledge that $\varphi_h+\varphi_l$ should be equal to $2\pi$. It is possible however, to construct a nearly perfectly optimized coating analytically and we will show that the ``common knowledge'' is in fact incorrect.

\begin{figure}[b]
\center
\includegraphics[width=0.4\textwidth]{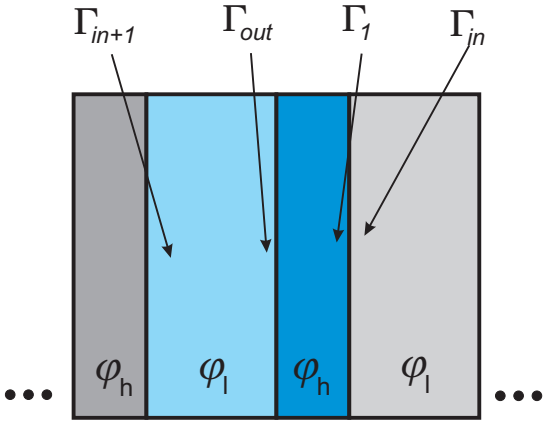}
\caption{A bilayer inside a multilayer coating}
\label{ilopt}
\end{figure}

We would like to find optimal thicknesses of the components of a bilayer for a given thickness $\varphi_h$. Suppose we have a bilayer inside a coating and the amplitude reflectivity on the boundary to this bilayer from the side of the substrate is $\Gamma_{in}=\Gamma_{0} e^{i\varphi_{0}}$, where $\Gamma_0$ is real amplitude and $\varphi_{0}$ is some initial phase. Let's introduce notations: $\Gamma_{in}=\Gamma_{0'}$ - initial reflectivity, $\Gamma_{1}$ - intermediate reflectivity, $\Gamma_{out}=\Gamma_{2}$ - output reflectivity, $\Gamma_{in+1}=\Gamma_{2'}$ - reflectivity, that will be initial for the next pair (see Fig. \ref{ilopt}):
\begin{align}
\Gamma_1=\frac{g_{hl}+\Gamma_{0'}}{1+g_{hl}\Gamma_{0'}}\\
\Gamma_2=\frac{g_{lh}+\Gamma_{1}e^{-i\varphi_h}}{1+g_{lh}\Gamma_{1}e^{-i\varphi_h}}\\
\Gamma_{2'}=\Gamma_2e^{-i\varphi_l},
\end{align}
where $g_{ij}=\frac{n_i-n_j}{n_i+n_j}$. Now we can find the optimal phase $\varphi_{0}$ maximizing $|\Gamma_{in+1}|^2$. Note that $\frac{\partial|\Gamma_{in+1}|^2}{\partial\varphi_{0}}=\frac{\partial|\Gamma_{out}|^2}{\partial\varphi_{0}}$ does not depend on $\varphi_l$. After some math we find
\begin{align}
\tan\varphi_{0}=\frac{1-g_{hl}^2}{1+g_{hl}^2}\cot\frac{\varphi_h}{2},\\
\label{phiin}
\varphi_0\approx\frac{\pi-\varphi_h}{2}-g^2\sin(\varphi_h).
\end{align}
In the last approximation we used the fact that $g_{hl}\simeq 0.17$ is small. It is also important that reflectivity increases with new pairs of layers only if $\varphi_0\in[-\frac{\pi}{2};\frac{\pi}{2}]$ (will be explained later).
To optimize the next layer, we should provide the same phase $\varphi_{in}$ for it, which gives
\begin{align}
\label{philuniv}
\varphi_{l_j}&=\varphi_{0_{j+1}}-\varphi_{0_j}-\varphi_{h_j}-2\sin(\varphi_{h_j})g_{hl_j}^2+(\pi m),
\end{align}
where $j$ stands for the bilayer number, $m$ is integer number. For a series of identical bilayers that means
\begin{align} 
\varphi_{l}&=-\varphi_h-2\sin(\varphi_h)g^2+(\pi m)
\end{align}
It can be shown that in our case ($\varphi_0\in[-\frac{\pi}{2};\frac{\pi}{2}]$, $\varphi_h\in[0;2\pi]$) $m>0$ and even. As we need to shorten layers for minimum noise $n=2$. Now from the last equation it is clear that only in the case 
of QWL coating $\varphi_H+\varphi_L=2\pi$. In other cases, however, there should be a small correction to maximize the reflection. Note that for the first layer $\Gamma_{in}=0$ with indefinite phase, thus satisfying the requirement on $\varphi_{0}$.

To make analytical approach consider a bilayer somewhere in the middle of coating. Suppose that incoming reflectivity is close to 1:
\begin{align}
\left|\Gamma_{in}\right|=\left|\Gamma_{2k'}\right|=1-\epsilon
\end{align}
Then, expanding formulas for $\left|\Gamma_{out}\right|$ into series to the second order of $\epsilon$ and using \eqref{phiin} we obtain
\begin{align}
\left|\Gamma_{out}\right|^2=1-2\alpha\epsilon-\alpha(1-2\alpha)\epsilon^2,
\end{align}
where
\begin{align}
\alpha&=\frac{(1-g^2)^2} {\left(g\sqrt{2(1-\cos(\varphi_1))}\pm\sqrt{1-2g^2\cos(\varphi_1)+g^4}\right)^2}
\end{align}
Here we have ``$+$'' sign, when $\varphi_0\in[-\pi/2;\pi/2]$. It can be shown, that only in this case $\alpha\leq1$, meaning increasing reflectivity.

Assuming $|2\alpha\epsilon|\ll 1$, $|(1-2\alpha)\epsilon|\ll 2$ we can rewrite local reflectivity as $\left|\Gamma_{out}\right|=1-\alpha\epsilon$ and get total power transmittance in the following form:
\begin{align}
\label{transmittance}
\tau=\beta(\varphi_h,\varphi_c)\alpha^N(\varphi_h)\epsilon(\varphi_e)
\end{align}
where
\begin{align}
\beta=2\frac{1-g_e^2}{1+2g_e\cos(\varphi_c+\frac{\pi+\varphi_h}{2}+g^2\sin(\varphi_h))+g_e^2}
\end{align}
describes the coating-air boundary ($g_{e}=\frac{n_e-n_l}{n_e+n_l}$).
For this formula to work we need to satisfy the assumptions we made. Calculations for $g=0.17$ give $\alpha\in[0.55;1]$ ($\alpha(\pi)=0.55$) and $\epsilon\ll0.5$. That also requires $\varphi_h\in[\pi/4;7\pi/4]$. It can be shown numerically that all those requirements can be satisfied with just three layers on the substrate.

Now we can eliminate the total number of layers from equations to design an optimal mirror for a given power transmittance. As calculations using \eqref{mainFormulas} - \eqref{mainKoefs} for the total noise are rather complicated but provide small correction only, we consider the simplified formula \eqref{BrownNoiseTrad}. For the spectral density we obtain
\begin{align}
S_{Br}=A\left((N+E)(\gamma \varphi_h+\varphi_l)+\varphi_c-\varphi_l\right)
\end{align}
in units of $\text{m}^2/\text{Hz}$ where $A=\frac{\xi_2\varphi_l}{2k_0n_l}$ is dimensional constant. Then for minimization we have
\begin{align}
\label{dimlessnoise}
\frac{S_{Br}}{A}=&E\gamma\varphi_{\epsilon h}+E\varphi_{\epsilon l}-\varphi_{0\epsilon}+\varphi_0+\\
	&\frac{\ln T_0-\ln\beta-\ln\epsilon}{\ln\alpha}(\gamma\varphi_h+\varphi_l)-\varphi_l+\varphi_c
\end{align}
Here $\varphi_{\epsilon h}$ $\varphi_{\epsilon l}$ denote phase thicknesses of first $E$ bilayers (from substrate), $\varphi_{0\epsilon}$, $\varphi_0$ - initial phases for initial (first $E$ layers) and regular bilayers, $\varphi_h$, $\varphi_l$ - regular layer thicknesses, $\varphi_c$ - cap layer thickness. 

All values here may be expressed in terms of $\varphi_h$, $\varphi_{\epsilon h}$, $\varphi_{c}$ and minimized. The results obtained are very close to \cite{Akira} (see table \ref{OptTab}) and represent an analytical alternative to numerical optimization.

\begin{table}
\begin{center}
\begin{tabular}{|c|c|c|c|}
\hline
Type & $25+\lambda/2$ & \cite{Akira} & Our Method \\\hline
Transmittance $\tau$, ppm & $277.5$ & $277.7$ & $277.7$  \\\hline
\hline
$\chi$, Brownian (displacement)	& $0$ & $8.16\%$& $8.4\%$  \\\hline
$\chi$, With interference & $3.37\%$ & $11.03\%$ & $11.27\%$ \\\hline
$\chi$, With photoelasticity& $2.63\%$ & $10.29\%$ & $10.5\%$ \\\hline
\hline
Real suppression & $0$ & $7.93\%$ & $8.18\%$ \\\hline
\end{tabular}
\caption{\label{OptTab}Optimization results. ($\gamma=7$)}
\end{center}

\end{table}

In conclusion, we found explicit formulas for spectral density of phase noise produced by Brownian thermal fluctuations in arbitrary multilayer coatings taking into account interference effects and photoelasticity in the coating. Interference effects and photoelasticity play a role only in few top layers and give correction of the order of $2\%$. Some optimization methods taking into account inerference were considered. The method of modifying silica-tantala ratio was found to be the most efficient so far. Another promising approach is compound mirrors.

\begin{acknowledgments}
M.L.G. acknowledges support from the Dynasty foundation, NSF grant PHY-0651036 and grant
08-02-00580 from Russian Foundation for Basic Research. Authors are grateful to S. P. Vyatchanin for stimulating discussions and to A. Villar for helpful remarks.
\end{acknowledgments}

\nocite{*}
\bibliographystyle{apsrev4-1}
\bibliography{intnoise}

\begin{thebibliography}{22}%
\makeatletter
\providecommand \@ifxundefined [1]{%
 \@ifx{#1\undefined}
}%
\providecommand \@ifnum [1]{%
 \ifnum #1\expandafter \@firstoftwo
 \else \expandafter \@secondoftwo
 \fi
}%
\providecommand \@ifx [1]{%
 \ifx #1\expandafter \@firstoftwo
 \else \expandafter \@secondoftwo
 \fi
}%
\providecommand \natexlab [1]{#1}%
\providecommand \enquote  [1]{``#1''}%
\providecommand \bibnamefont  [1]{#1}%
\providecommand \bibfnamefont [1]{#1}%
\providecommand \citenamefont [1]{#1}%
\providecommand \href@noop [0]{\@secondoftwo}%
\providecommand \href [0]{\begingroup \@sanitize@url \@href}%
\providecommand \@href[1]{\@@startlink{#1}\@@href}%
\providecommand \@@href[1]{\endgroup#1\@@endlink}%
\providecommand \@sanitize@url [0]{\catcode `\\12\catcode `\$12\catcode
  `\&12\catcode `\#12\catcode `\^12\catcode `\_12\catcode `\%12\relax}%
\providecommand \@@startlink[1]{}%
\providecommand \@@endlink[0]{}%
\providecommand \url  [0]{\begingroup\@sanitize@url \@url }%
\providecommand \@url [1]{\endgroup\@href {#1}{\urlprefix }}%
\providecommand \urlprefix  [0]{URL }%
\providecommand \Eprint [0]{\href }%
\providecommand \doibase [0]{http://dx.doi.org/}%
\providecommand \selectlanguage [0]{\@gobble}%
\providecommand \bibinfo  [0]{\@secondoftwo}%
\providecommand \bibfield  [0]{\@secondoftwo}%
\providecommand \translation [1]{[#1]}%
\providecommand \BibitemOpen [0]{}%
\providecommand \bibitemStop [0]{}%
\providecommand \bibitemNoStop [0]{.\EOS\space}%
\providecommand \EOS [0]{\spacefactor3000\relax}%
\providecommand \BibitemShut  [1]{\csname bibitem#1\endcsname}%
\let\auto@bib@innerbib\@empty
\bibitem [{\citenamefont {Abbott}\ \emph {et~al.}(2009)\citenamefont {Abbott}
  \emph {et~al.}}]{LIGO}%
  \BibitemOpen
  \bibfield  {author} {\bibinfo {author} {\bibfnamefont {B.}~\bibnamefont
  {Abbott}} \emph {et~al.},\ }\href@noop {} {\bibfield  {journal} {\bibinfo
  {journal} {Rep. Progr. Phys.}\ }\textbf {\bibinfo {volume} {72}},\ \bibinfo
  {pages} {076901:25} (\bibinfo {year} {2009})}\BibitemShut {NoStop}%
\bibitem [{\citenamefont {Gorodetsky}(2008)}]{Zoo}%
  \BibitemOpen
  \bibfield  {author} {\bibinfo {author} {\bibfnamefont {M.~L.}\ \bibnamefont
  {Gorodetsky}},\ }\href@noop {} {\bibfield  {journal} {\bibinfo  {journal}
  {Phys. Lett. A}\ }\textbf {\bibinfo {volume} {372}},\ \bibinfo {pages} {6813}
  (\bibinfo {year} {2008})}\BibitemShut {NoStop}%
\bibitem [{\citenamefont {Gurkovsky}\ and\ \citenamefont
  {Vyatchanin}(2010)}]{Gurkovsky}%
  \BibitemOpen
  \bibfield  {author} {\bibinfo {author} {\bibfnamefont {A.}~\bibnamefont
  {Gurkovsky}}\ and\ \bibinfo {author} {\bibfnamefont {S.}~\bibnamefont
  {Vyatchanin}},\ }\href@noop {} {\bibfield  {journal} {\bibinfo  {journal}
  {Phys. Lett. A}\ }\textbf {\bibinfo {volume} {374}},\ \bibinfo {pages} {3267}
  (\bibinfo {year} {2010})}\BibitemShut {NoStop}%
\bibitem [{\citenamefont {Villar}\ \emph
  {et~al.}(2010{\natexlab{a}})\citenamefont {Villar}, \citenamefont {Black},
  \citenamefont {DeSalvo}, \citenamefont {Libbrecht}, \citenamefont {Michel},
  \citenamefont {Morgado}, \citenamefont {Pinard}, \citenamefont {Pinto},
  \citenamefont {Pierro}, \citenamefont {Galdi}, \citenamefont {Principe},\
  and\ \citenamefont {Taurasi}}]{Akira}%
  \BibitemOpen
  \bibfield  {author} {\bibinfo {author} {\bibfnamefont {A.~E.}\ \bibnamefont
  {Villar}}, \bibinfo {author} {\bibfnamefont {E.~D.}\ \bibnamefont {Black}},
  \bibinfo {author} {\bibfnamefont {R.}~\bibnamefont {DeSalvo}}, \bibinfo
  {author} {\bibfnamefont {K.~G.}\ \bibnamefont {Libbrecht}}, \bibinfo {author}
  {\bibfnamefont {C.}~\bibnamefont {Michel}}, \bibinfo {author} {\bibfnamefont
  {N.}~\bibnamefont {Morgado}}, \bibinfo {author} {\bibfnamefont
  {L.}~\bibnamefont {Pinard}}, \bibinfo {author} {\bibfnamefont {I.~M.}\
  \bibnamefont {Pinto}}, \bibinfo {author} {\bibfnamefont {V.}~\bibnamefont
  {Pierro}}, \bibinfo {author} {\bibfnamefont {V.}~\bibnamefont {Galdi}},
  \bibinfo {author} {\bibfnamefont {M.}~\bibnamefont {Principe}}, \ and\
  \bibinfo {author} {\bibfnamefont {I.}~\bibnamefont {Taurasi}},\ }\href@noop
  {} {\bibfield  {journal} {\bibinfo  {journal} {Phys. Rev. D}\ }\textbf
  {\bibinfo {volume} {81}},\ \bibinfo {pages} {122001} (\bibinfo {year}
  {2010}{\natexlab{a}})}\BibitemShut {NoStop}%
\bibitem [{\citenamefont {Harry}\ \emph {et~al.}(2002)\citenamefont {Harry},
  \citenamefont {Gretarsson}, \citenamefont {Saulson}, \citenamefont {Penn},
  \citenamefont {Startin}, \citenamefont {Rowan}, \citenamefont {Fejer},
  \citenamefont {Crooks}, \citenamefont {Cagnoli}, \citenamefont {Hough},\ and\
  \citenamefont {Nakagawa}}]{Harry}%
  \BibitemOpen
  \bibfield  {author} {\bibinfo {author} {\bibfnamefont {G.~M.}\ \bibnamefont
  {Harry}}, \bibinfo {author} {\bibfnamefont {A.~M.}\ \bibnamefont
  {Gretarsson}}, \bibinfo {author} {\bibfnamefont {S.~E.}\ \bibnamefont
  {Saulson}, \bibfnamefont {P.~R.~Kittelberger}}, \bibinfo {author}
  {\bibfnamefont {S.~D.}\ \bibnamefont {Penn}}, \bibinfo {author}
  {\bibfnamefont {W.~J.}\ \bibnamefont {Startin}}, \bibinfo {author}
  {\bibfnamefont {S.}~\bibnamefont {Rowan}}, \bibinfo {author} {\bibfnamefont
  {M.~M.}\ \bibnamefont {Fejer}}, \bibinfo {author} {\bibfnamefont {D.~R.~M.}\
  \bibnamefont {Crooks}}, \bibinfo {author} {\bibfnamefont {G.}~\bibnamefont
  {Cagnoli}}, \bibinfo {author} {\bibfnamefont {J.}~\bibnamefont {Hough}}, \
  and\ \bibinfo {author} {\bibfnamefont {N.}~\bibnamefont {Nakagawa}},\
  }\href@noop {} {\bibfield  {journal} {\bibinfo  {journal} {Clas. Quantum
  Grav.}\ }\textbf {\bibinfo {volume} {19}},\ \bibinfo {pages} {897} (\bibinfo
  {year} {2002})}\BibitemShut {NoStop}%
\bibitem [{\citenamefont {Kimble}\ \emph {et~al.}(2008)\citenamefont {Kimble},
  \citenamefont {Lev},\ and\ \citenamefont {Ye}}]{Kimble}%
  \BibitemOpen
  \bibfield  {author} {\bibinfo {author} {\bibfnamefont {H.~J.}\ \bibnamefont
  {Kimble}}, \bibinfo {author} {\bibfnamefont {B.~L.}\ \bibnamefont {Lev}}, \
  and\ \bibinfo {author} {\bibfnamefont {J.}~\bibnamefont {Ye}},\ }\href@noop
  {} {\bibfield  {journal} {\bibinfo  {journal} {Phys. Rev. Lett.}\ }\textbf
  {\bibinfo {volume} {101}},\ \bibinfo {pages} {260602} (\bibinfo {year}
  {2008})}\BibitemShut {NoStop}%
\bibitem [{\citenamefont {Braginsky}\ \emph {et~al.}(1999)\citenamefont
  {Braginsky}, \citenamefont {Gorodetsky},\ and\ \citenamefont
  {Vyatchanin}}]{BGVte}%
  \BibitemOpen
  \bibfield  {author} {\bibinfo {author} {\bibfnamefont {V.~B.}\ \bibnamefont
  {Braginsky}}, \bibinfo {author} {\bibfnamefont {M.~L.}\ \bibnamefont
  {Gorodetsky}}, \ and\ \bibinfo {author} {\bibfnamefont {S.~P.}\ \bibnamefont
  {Vyatchanin}},\ }\href@noop {} {\bibfield  {journal} {\bibinfo  {journal}
  {Phys. Lett. A}\ }\textbf {\bibinfo {volume} {264}},\ \bibinfo {pages} {1}
  (\bibinfo {year} {1999})}\BibitemShut {NoStop}%
\bibitem [{\citenamefont {Braginsky}\ and\ \citenamefont
  {Vyatchanin}(2003)}]{BGVteC}%
  \BibitemOpen
  \bibfield  {author} {\bibinfo {author} {\bibfnamefont {V.~B.}\ \bibnamefont
  {Braginsky}}\ and\ \bibinfo {author} {\bibfnamefont {S.~P.}\ \bibnamefont
  {Vyatchanin}},\ }\href@noop {} {\bibfield  {journal} {\bibinfo  {journal}
  {Phys. Lett. A}\ }\textbf {\bibinfo {volume} {312}},\ \bibinfo {pages} {244}
  (\bibinfo {year} {2003})}\BibitemShut {NoStop}%
\bibitem [{\citenamefont {Braginsky}\ \emph {et~al.}(2000)\citenamefont
  {Braginsky}, \citenamefont {Gorodetsky},\ and\ \citenamefont
  {Vyatchanin}}]{BGVtr}%
  \BibitemOpen
  \bibfield  {author} {\bibinfo {author} {\bibfnamefont {V.~B.}\ \bibnamefont
  {Braginsky}}, \bibinfo {author} {\bibfnamefont {M.~L.}\ \bibnamefont
  {Gorodetsky}}, \ and\ \bibinfo {author} {\bibfnamefont {S.~P.}\ \bibnamefont
  {Vyatchanin}},\ }\href@noop {} {\bibfield  {journal} {\bibinfo  {journal}
  {Phys. Lett. A}\ }\textbf {\bibinfo {volume} {271}},\ \bibinfo {pages} {303}
  (\bibinfo {year} {2000})}\BibitemShut {NoStop}%
\bibitem [{\citenamefont {Evans}\ \emph {et~al.}(2008)\citenamefont {Evans},
  \citenamefont {Ballmer}, \citenamefont {Fejer}, \citenamefont {Fritschel},
  \citenamefont {Harry},\ and\ \citenamefont {Ogin}}]{Evans}%
  \BibitemOpen
  \bibfield  {author} {\bibinfo {author} {\bibfnamefont {M.}~\bibnamefont
  {Evans}}, \bibinfo {author} {\bibfnamefont {S.}~\bibnamefont {Ballmer}},
  \bibinfo {author} {\bibfnamefont {M.}~\bibnamefont {Fejer}}, \bibinfo
  {author} {\bibfnamefont {P.}~\bibnamefont {Fritschel}}, \bibinfo {author}
  {\bibfnamefont {G.}~\bibnamefont {Harry}}, \ and\ \bibinfo {author}
  {\bibfnamefont {G.}~\bibnamefont {Ogin}},\ }\href@noop {} {\bibfield
  {journal} {\bibinfo  {journal} {Phys. Rev. D}\ }\textbf {\bibinfo {volume}
  {78}},\ \bibinfo {pages} {102003} (\bibinfo {year} {2008})}\BibitemShut
  {NoStop}%
\bibitem [{\citenamefont {Haus}(1993)}]{Haus}%
  \BibitemOpen
  \bibfield  {author} {\bibinfo {author} {\bibfnamefont {H.~A.}\ \bibnamefont
  {Haus}},\ }\href@noop {} {\emph {\bibinfo {title} {Waves and Fields in
  Optoelectronics}}}\ (\bibinfo  {publisher} {Prentice Hall},\ \bibinfo {year}
  {1993})\BibitemShut {NoStop}%
\bibitem [{\citenamefont {Furman}\ and\ \citenamefont
  {Tikhonravov}(1992)}]{Tikhonravov}%
  \BibitemOpen
  \bibfield  {author} {\bibinfo {author} {\bibfnamefont {S.~A.}\ \bibnamefont
  {Furman}}\ and\ \bibinfo {author} {\bibfnamefont {A.~V.}\ \bibnamefont
  {Tikhonravov}},\ }\href@noop {} {\emph {\bibinfo {title} {Basics of Optics of
  Multilayer Systems}}}\ (\bibinfo  {publisher} {Atlantica S\'eguier
  Fronti\`eres},\ \bibinfo {year} {1992})\BibitemShut {NoStop}%
\bibitem [{\citenamefont {Yariv}\ and\ \citenamefont {Yeh}(2003)}]{Acoustoopt}%
  \BibitemOpen
  \bibfield  {author} {\bibinfo {author} {\bibfnamefont {A.}~\bibnamefont
  {Yariv}}\ and\ \bibinfo {author} {\bibfnamefont {P.}~\bibnamefont {Yeh}},\
  }\href@noop {} {\emph {\bibinfo {title} {Optical Waves in Crystals:
  Propagation and Control of Laser Radiation}}}\ (\bibinfo  {publisher}
  {Wiley-Interscience},\ \bibinfo {year} {2003})\BibitemShut {NoStop}%
\bibitem [{\citenamefont {Wille}\ and\ \citenamefont
  {Hamilton}(1974)}]{Tantal}%
  \BibitemOpen
  \bibfield  {author} {\bibinfo {author} {\bibfnamefont {D.~A.}\ \bibnamefont
  {Wille}}\ and\ \bibinfo {author} {\bibfnamefont {M.~C.}\ \bibnamefont
  {Hamilton}},\ }\href@noop {} {\bibfield  {journal} {\bibinfo  {journal}
  {Appl. Phys. Lett.}\ }\textbf {\bibinfo {volume} {{24}}},\ \bibinfo {pages}
  {{159}} (\bibinfo {year} {{1974}})}\BibitemShut {NoStop}%
\bibitem [{\citenamefont {Khalili}(2005)}]{Khalili}%
  \BibitemOpen
  \bibfield  {author} {\bibinfo {author} {\bibfnamefont {F.~Y.}\ \bibnamefont
  {Khalili}},\ }\href@noop {} {\bibfield  {journal} {\bibinfo  {journal} {Phys.
  Lett. A}\ }\textbf {\bibinfo {volume} {334}},\ \bibinfo {pages} {67}
  (\bibinfo {year} {2005})}\BibitemShut {NoStop}%
\bibitem [{\citenamefont {Agresti}\ \emph {et~al.}(2006)\citenamefont
  {Agresti}, \citenamefont {Castaldi}, \citenamefont {DeSalvo}, \citenamefont
  {Galdi}, \citenamefont {Pierro},\ and\ \citenamefont {Pinto}}]{Pinto}%
  \BibitemOpen
  \bibfield  {author} {\bibinfo {author} {\bibfnamefont {J.}~\bibnamefont
  {Agresti}}, \bibinfo {author} {\bibfnamefont {G.}~\bibnamefont {Castaldi}},
  \bibinfo {author} {\bibfnamefont {R.}~\bibnamefont {DeSalvo}}, \bibinfo
  {author} {\bibfnamefont {V.}~\bibnamefont {Galdi}}, \bibinfo {author}
  {\bibfnamefont {V.}~\bibnamefont {Pierro}}, \ and\ \bibinfo {author}
  {\bibfnamefont {I.~M.}\ \bibnamefont {Pinto}},\ }in\ \href@noop {} {\emph
  {\bibinfo {booktitle} {{Proc. SPIE}}}},\ Vol.\ \bibinfo {volume} {6286}\
  (\bibinfo {year} {2006})\ p.\ \bibinfo {pages} {28608}\BibitemShut {NoStop}%
\bibitem [{GWI()}]{GWINC}%
  \BibitemOpen
  \href@noop {} {\enquote {\bibinfo {title} {Gravitational wave interferometer
  noise calculator ({GWINC})},}\ }\bibinfo {note}
  {{http://ilog.ligo.org/cgi-bin/DocDB/RetrieveFile?docid=1507}}\BibitemShut
  {NoStop}%
\bibitem [{\citenamefont {Villar}\ \emph
  {et~al.}(2010{\natexlab{b}})\citenamefont {Villar}, \citenamefont {Black},
  \citenamefont {Ogin}, \citenamefont {Chelermsongsak}, \citenamefont
  {DeSalvo}, \citenamefont {Pinto}, \citenamefont {Pierro},\ and\ \citenamefont
  {Principe}}]{Krakow}%
  \BibitemOpen
  \bibfield  {author} {\bibinfo {author} {\bibfnamefont {A.~V.}\ \bibnamefont
  {Villar}}, \bibinfo {author} {\bibfnamefont {E.}~\bibnamefont {Black}},
  \bibinfo {author} {\bibfnamefont {G.}~\bibnamefont {Ogin}}, \bibinfo {author}
  {\bibfnamefont {T.}~\bibnamefont {Chelermsongsak}}, \bibinfo {author}
  {\bibfnamefont {R.}~\bibnamefont {DeSalvo}}, \bibinfo {author} {\bibfnamefont
  {I.}~\bibnamefont {Pinto}}, \bibinfo {author} {\bibfnamefont
  {V.}~\bibnamefont {Pierro}}, \ and\ \bibinfo {author} {\bibfnamefont
  {M.}~\bibnamefont {Principe}},\ }in\ \href@noop {} {\emph {\bibinfo
  {booktitle} {LSC-Virgo meeting in Krakow}}},\ \bibinfo {series and number}
  {\bibinfo {number} {G1000937}}\ (\bibinfo {year} {{2010}})\BibitemShut
  {NoStop}%
\bibitem [{\citenamefont {Levin}(2008)}]{Levin}%
  \BibitemOpen
  \bibfield  {author} {\bibinfo {author} {\bibfnamefont {Y.}~\bibnamefont
  {Levin}},\ }\href@noop {} {\bibfield  {journal} {\bibinfo  {journal} {Phys.
  Lett. A}\ }\textbf {\bibinfo {volume} {372}},\ \bibinfo {pages} {1941}
  (\bibinfo {year} {2008})}\BibitemShut {NoStop}%
\bibitem [{\citenamefont {Levin}(1998)}]{LevinB}%
  \BibitemOpen
  \bibfield  {author} {\bibinfo {author} {\bibfnamefont {Y.}~\bibnamefont
  {Levin}},\ }\href@noop {} {\bibfield  {journal} {\bibinfo  {journal} {Phys.
  Rev. D}\ }\textbf {\bibinfo {volume} {57}},\ \bibinfo {pages} {659} (\bibinfo
  {year} {1998})}\BibitemShut {NoStop}%
\bibitem [{\citenamefont {Harry}\ \emph {et~al.}(2006)\citenamefont {Harry},
  \citenamefont {Armandula}, \citenamefont {Black}, \citenamefont {Crooks},
  \citenamefont {Cagnoli}, \citenamefont {Hough}, \citenamefont {Murray},
  \citenamefont {Reid}, \citenamefont {Rowan}, \citenamefont {Sneddon},
  \citenamefont {Fejer}, \citenamefont {Route},\ and\ \citenamefont
  {Penn}}]{HarryB}%
  \BibitemOpen
  \bibfield  {author} {\bibinfo {author} {\bibfnamefont {G.~M.}\ \bibnamefont
  {Harry}}, \bibinfo {author} {\bibfnamefont {H.}~\bibnamefont {Armandula}},
  \bibinfo {author} {\bibfnamefont {E.}~\bibnamefont {Black}}, \bibinfo
  {author} {\bibfnamefont {D.~R.~M.}\ \bibnamefont {Crooks}}, \bibinfo {author}
  {\bibfnamefont {G.}~\bibnamefont {Cagnoli}}, \bibinfo {author} {\bibfnamefont
  {J.}~\bibnamefont {Hough}}, \bibinfo {author} {\bibfnamefont
  {P.}~\bibnamefont {Murray}}, \bibinfo {author} {\bibfnamefont
  {S.}~\bibnamefont {Reid}}, \bibinfo {author} {\bibfnamefont {S.}~\bibnamefont
  {Rowan}}, \bibinfo {author} {\bibfnamefont {P.}~\bibnamefont {Sneddon}},
  \bibinfo {author} {\bibfnamefont {M.~M.}\ \bibnamefont {Fejer}}, \bibinfo
  {author} {\bibfnamefont {R.}~\bibnamefont {Route}}, \ and\ \bibinfo {author}
  {\bibfnamefont {S.~D.}\ \bibnamefont {Penn}},\ }\href@noop {} {\bibfield
  {journal} {\bibinfo  {journal} {Appl. Opt.}\ }\textbf {\bibinfo {volume}
  {45}},\ \bibinfo {pages} {1569} (\bibinfo {year} {2006})}\BibitemShut
  {NoStop}%
\bibitem [{\citenamefont {Somiya}\ \emph {et~al.}(2010)\citenamefont {Somiya},
  \citenamefont {Gurkovsky}, \citenamefont {Vyatchanin}, \citenamefont
  {Heinert}, \citenamefont {Nawrodt},\ and\ \citenamefont {Hild}}]{Kentaro}%
  \BibitemOpen
  \bibfield  {author} {\bibinfo {author} {\bibfnamefont {K.}~\bibnamefont
  {Somiya}}, \bibinfo {author} {\bibfnamefont {A.~G.}\ \bibnamefont
  {Gurkovsky}}, \bibinfo {author} {\bibfnamefont {S.~P.}\ \bibnamefont
  {Vyatchanin}}, \bibinfo {author} {\bibfnamefont {D.}~\bibnamefont {Heinert}},
  \bibinfo {author} {\bibfnamefont {R.}~\bibnamefont {Nawrodt}}, \ and\
  \bibinfo {author} {\bibfnamefont {S.}~\bibnamefont {Hild}},\ }\href@noop {}
  {\emph {\bibinfo {title} {Reduction of coating thermal noise by using an
  etalon}}},\ \bibinfo {type} {Tech. Rep.}\ \bibinfo {number} {P1000094-v1}\
  (\bibinfo  {institution} {LIGO},\ \bibinfo {year} {{2010}})\BibitemShut
  {NoStop}%
\end{thebibliography}%

\end{document}